 \documentclass{ws-procs9x6}
\begin{document}

\title{Path Integrals in Lattice Quantum Chromodynamics}
\author{Frank X. Lee~\footnote{E-mail: fxlee@gwu.edu}}
\address{Physics Department, George Washington University, Washington, DC 20052, USA}

\begin{abstract}
I discuss the use of path integrals to study strong-interaction physics from first principles. 
The underlying theory is cast into path integrals which are evaluated numerically 
using Monte Carlo methods on a space-time lattice. 
Examples are given on progress related to nuclear physics.
\end{abstract}


\bodymatter

\section{Introduction}
Quantum chromodynamics (or QCD) is the underlying theory of the strong interaction 
(one of the four fundamental interactions in nature besides gravity, electromagnetism
and the weak interaction). It is the force that binds quarks and gluons as the nucleus
in the heart of the atom.  Since all matter in the universe is built that way, 
unraveling the structure of matter at its deepest level as governed by QCD is key to our 
understanding of the physical world, 
and presents one of the most challenging tasks in contemporary nuclear and particle physics.  
While it is easy to write down the basic equation of QCD, 
it is very difficult to obtain quantitative solutions in view of the complex quark-gluon 
dynamics. 
At present, the only known way to solve QCD directly is by numerically evaluating 
the path integrals in the theory on a discrete space-time lattice 
using supercomputers. For textbooks on the subject, see Refs.~\cite{rothe,montvay,degrand}.
%
%

\section{Formalism}
All physics can be computed by path integrals in QCD. 
Take the calculation of the proton mass for example. It requires 
the fully-interacting quark propagator defined by the Euclidean-space path integrals over 
gluon field $A_\mu$ and quarks fields $\psi$ and $\bar{\psi}$,
\begin{equation}
<M^{-1}>\equiv <0|\psi(x)\bar{\psi}(0)|0>\equiv 
\frac{\int D\psi D\bar{\psi}DA_\mu  e^{-S_{QCD}} M^{-1} }
{\int D\psi D\bar{\psi}DA_\mu  e^{-S_{QCD}} }.
\end{equation}
The action is the sum of a gluon part and a quark part 
\begin{equation}
S_{QCD}=S_G + S_q =
\frac{1}{2} \int dx^4  \mbox{Tr} F_{\mu\nu}\,F^{\mu\nu}
+ \int dx^4 \bar{\psi} M \psi,
\end{equation}
where $F_{\mu\nu}=\partial_\mu A_\nu- \partial_\nu A_\mu + g[A_\mu,A_\nu]$ 
is the gluon field strength tensor and 
$M=\gamma^\mu D_\mu + m_q$ is the Dirac operator for quarks and $\gamma^\mu$ the 
4x4 gamma matrices.
The interaction between the two is provided by the covariant derivative 
$D_\mu=\partial_\mu+gA_\mu$. The basic parameters of the theory are 
the coupling constant $g$ and quark mass $m_q$.
The quark part of the integration can be done analytically using Grassmann variable integration, 
leading to a path integral over only gluon fields
\begin{equation}
<M^{-1}>\equiv 
\frac{\int DA_\mu  \mbox{det}(M) e^{-S_G} M^{-1} }
{\int DA_\mu  \mbox{det}(M) e^{-S_G}  }
\label{qprop}
\end{equation}
which is evaluated numerically by Monte-Carlo methods.
In the quenched approximation, the quark determinant $\mbox{det}(M)$ is ignored (equivalent 
to suppressing vacuum polarization), 
making the numerical calculation significantly faster.

To compute the proton mass, one considers the time-ordered, two-point correlation
function in the QCD vacuum, projected to zero momentum:
\begin{equation}
G(t)=\sum_{\vec{x}}\langle 0\,|\, T\{\;\eta(x)\, \bar{\eta}(0)\;\}\,|\,0\rangle.
\label{cf2pt}
\end{equation}
Here $\eta$ is called the interpolating field which is built from quark fields
with the quantum numbers of the proton (spin-1/2, isospin-1/2, quark content uud, charge +e)
\begin{equation}
\eta(x)=\epsilon^{abc} \left[u^{aT}(x)C\gamma_5 d^b(x)\right]u^c(x)
\end{equation}
where $C$ is the charge conjugation operator and the superscript $T$ means transpose.  
Sum over color indeces a,b,c is implied and the $\epsilon^{abc}$ ensures the proton is 
color-singlet.

The calculation of $G(t)$ at the quark level proceeds by contracting out all the
quark pairs, resulting in
\begin{equation}
G(t) = \sum_{\vec{x}}
\epsilon^{abc}\epsilon^{a^\prime b^\prime c^\prime} 
\left\{
S_u^{aa^\prime} \gamma_5
C {S^{cc^\prime}_d}^T C  \gamma_5 S^{bb^\prime}_u
+S^{aa^\prime}_u \mbox{Tr}(C {S^{cc^\prime}}^T_d C
\gamma_5 S^{bb^\prime}_u \gamma_5) \right\},
\end{equation}
where $S_q^{ij}$ denotes the fully-interacting quark propagator $<M^{-1}>$ in Eq.~(\ref{qprop}).

On the hadronic level, the correlation function is saturated by the complete spectrum of 
intermediate states (after a parity projection) with the proton as the ground state:
\begin{equation}
G(t)=\sum_i \lambda^2_i\,e^{-m_i t}
\label{pole2}
\end{equation}
where $m_i$ are the masses and $\lambda^2$ are the 'amplitudes' 
which are a measure of the ability of the
interpolating field to excite or annhilate the states from the QCD vacuum.
At large time, the ground state proton, $\lambda^2_1\,e^{-m_1 t}$, dominates $G(t)$, 
with the excited states exponentially suppressed.
Other physics quantities are computed in a similar way.

\section{Some examples}
It is impossible to give a full account of the achievements made in lattice QCD 
in the space given here.
A good place to gauge the progress in the field (which is not limited to QCD) 
is the annual gathering~\cite{latxx} by active practitioners.
Here I select a few examples relevant to nuclear physics.

\begin{figure}
\vspace*{-5mm}
\centerline{\psfig{file=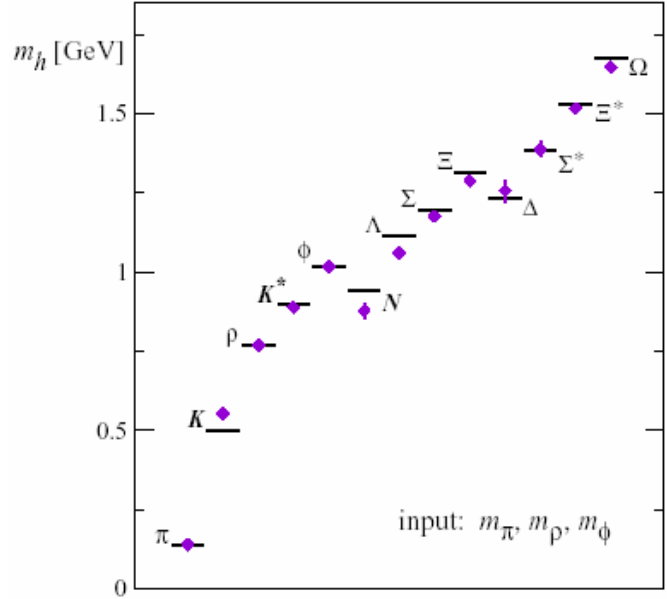,width=3.5in,angle=0}}
\caption{The computed light hadron mass spectrum in the quenched approximation.
See Ref.~\protect\cite{cppacs00} for details.}
\label{cppacs}
\end{figure}

{\bf Example 1}.
Fig.~\ref{cppacs} shows that the computed light hadron mass spectrum which comes within 10\% of 
the observed spectrum. The remaining discrepancy is attributed to the quenched approximation.
Intense efforts are under way to extend the success to the excited sectors of the mass
spectrum~\cite{lhpc07}.

\begin{figure}
\centerline{\psfig{file=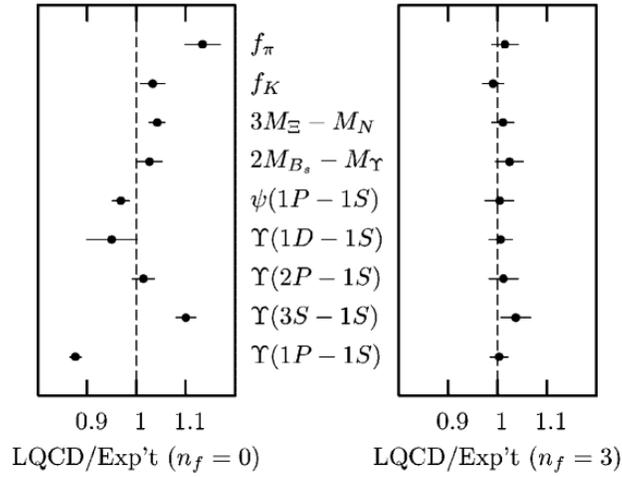,width=3.2in,angle=0}}
\caption{Lattice QCD results compared to experiment as
ratios for selected physics quantities (the dashed line is perfect agreement), 
in the quenched approximation of QCD (left panel) and full QCD (right panel).
See Ref.~\protect\cite{hpqcd04} for details.}
\label{hpqcd04}
\end{figure}

{\bf Example 2}.
Fig.~\ref{hpqcd04} shows the high-precision lattice QCD calculations for selected physics
quantities which are within 3\% of the observed values.  The effects of the quenched 
approximation are clearly demonstrated.

\begin{figure}
\centerline{\psfig{file=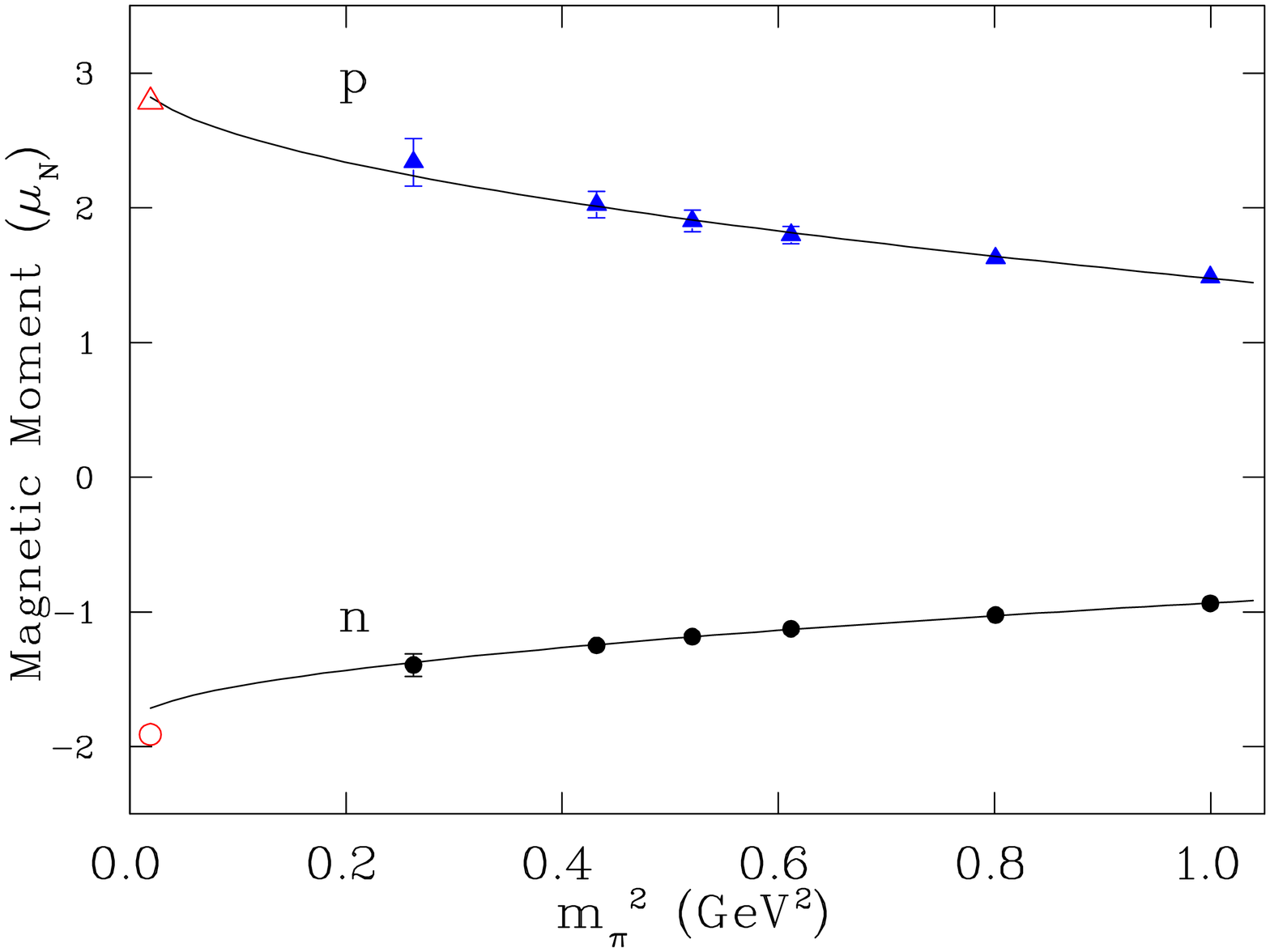,width=2.0in,angle=90}}
\caption{Magnetic moment of the proton and neutron as a function of the pion mass squared. 
The line is a chiral extrapolation.
See Ref.~\protect\cite{Lee05} for details.}
\label{mag}
\end{figure}

{\bf Example 3}.
Fig.~\ref{mag} shows a calculation of the magnetic moment of the proton ans neutron 
which find good agreement between lattice QCD and experiment. 
Results from a different method can be found in Ref.~\cite{Boin06}, 
as well as precise determinations of the strangeness magnetic ($G^s_M$)~\cite{Derek05} 
and electric form factors ($G^s_E$)~\cite{Derek06} of the proton.

\begin{figure}
\centerline{\psfig{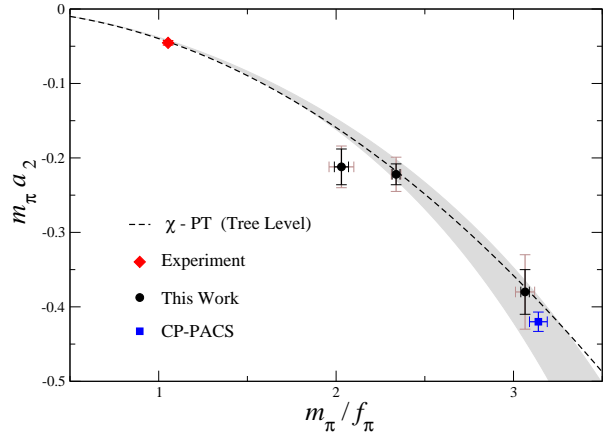}}
\caption{Lattice QCD calculation of the $\pi-\pi$ scattering length in the isospin 2 channel.
The dashed line is a chiral fit.
See Ref.~\protect\cite{pipi06} for details.}
\label{pipi}
\end{figure}

{\bf Example 4}.
Fig.~\ref{pipi} shows the good agreement between lattice QCD and experiment 
for pion-pion scattering length~\cite{pipi06}, using L\"{u}scher's method~\cite{Luch91}.
Other hadron-hadron scattering channels ($\pi^+K^+$, $n\Lambda$, $n\Sigma^-$, {\it etc.}) 
have been studied by the same group.

\begin{figure}
\centerline{\psfig{file=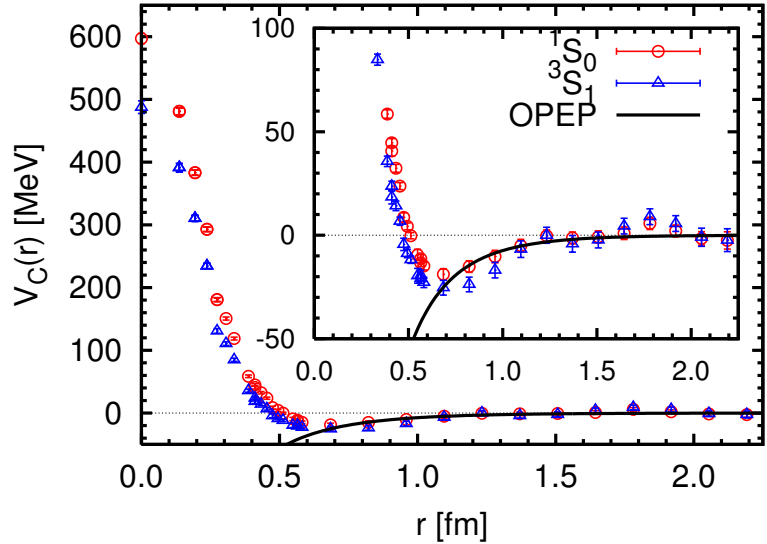,width=3.0in,angle=-90}}
\caption{The lattice QCD result of the central (effective central)
part of the NN potential in the $^1$S$_0$ ($^3$S$_1$) channels. The inset shows its
enlargement. The solid lines correspond to the one-pion exchange potential (Yukawa potential).
See Ref.~\protect\cite{aoki07} for details.}
\label{nucpot}
\end{figure}

{\bf Example 5}.
Fig.~\ref{nucpot} shows a direct lattice QCD calculation of the nucleon-nucleon 
potential~\cite{aoki07} whose features are
consistent with the known phenomenological features of the nuclear force.
%

\section{Conclusion}
The path-integral formulation of QCD, coupled with large-scale numerical simulations, 
has played a crucial role in the progress of nuclear and particle physics. 
Its continued prominence as the only way to solve QCD with controlled 
errors is expected to play out with better results in the foreseeable future.
This work is supported in part by U.S. Department of Energy under grant DE-FG02-95ER40907.

\end{document}